\documentclass[preprint]{aastex}
\usepackage{emulateapj5,natbib}

\slugcomment{\it Accepted for publication in the Astrophysical
Journal, June 17, 2003}

\shortauthors{McCRADY, GILBERT, \& GRAHAM}
\shorttitle{SSC MASSES IN M82}

\begin{document}

\title{KINEMATIC MASSES OF SUPER STAR CLUSTERS IN M82 
FROM HIGH-RESOLUTION NEAR-INFRARED SPECTROSCOPY\altaffilmark{1}
}

\author{
Nate McCrady\altaffilmark{2}, Andrea M. Gilbert and James R. Graham
}

\affil{
Department of Astronomy, University of California, Berkeley, CA 94720-3411
}

\altaffiltext{1}{
Based on observations made at the W.M. Keck Observatory, which is operated
as a scientific partnership among the California Institute of Technology,
the University of California and the National Aeronautics and Space
Administration.  The Observatory was made possible by the generous
financial support of the W.M. Keck Foundation.}

\altaffiltext{2}{
nate@astro.berkeley.edu}

%%%%%%%%%%%%
% Abstract %
%%%%%%%%%%%%

\begin{abstract}
Using high-resolution ($R \sim $ 22,000) near-infrared (1.51 -- 1.75
$\mu$m) spectra from Keck Observatory, we measure the kinematic masses
of two super star clusters (SSCs) in M82.  Cross-correlation of the
spectra with template spectra of cool evolved stars gives stellar
velocity dispersions of $\sigma_r = 15.9 \pm 0.8$ km s$^{-1}$ for
J0955505+694045 (`MGG-9') and $\sigma_r = 11.4 \pm 0.8$ km s$^{-1}$
for J0955502+694045 (`MGG-11').  The cluster spectra are dominated by
the light of red supergiants, and correlate most closely with template
supergiants of spectral types M0 and M4.5.  King model fits to the
observed profiles of the clusters in archival HST/NICMOS images give
half-light radii of $r_{hp} = 2.6 \pm 0.4$ pc for MGG-9 and $r_{hp} =
1.2 \pm 0.17$ pc for MGG-11.  Applying the virial theorem, we
determine masses of $1.5 \pm 0.3 \times 10^6$ M$_{\odot}$ for MGG-9
and $3.5 \pm 0.7 \times 10^5$ M$_{\odot}$ for MGG-11 (where the quoted
errors include $\sigma_r$, $r_{hp}$ and the distance).  Population
synthesis modelling suggests that MGG-9 is consistent with a standard
initial mass function, whereas MGG-11 appears to be deficient in
low-mass stars relative to a standard IMF.  There is, however,
evidence of mass segregation in the clusters, in which case the virial
mass estimates would represent lower limits.

\end{abstract}


%%%%%%%%
% Body %
%%%%%%%%

\section{Introduction}
\label{intro}

Starburst galaxies are the site of about 25\% of the high-mass ($>$
8M$_{\odot}$) star formation in the local Universe \citep{heckman98}.
As high-mass stars form predominantly in dense clusters and OB
associations \citep{miller78}, massive stellar clusters in nearby
starburst galaxies provide a laboratory for studying intense star
formation and related feedback processes.  Star formation in
starbursts is resolved into massive, compact super star clusters
(SSCs) which account for perhaps 20\% of the new stars formed in the
burst \citep{meurer95,zepf99}.  The combination of high-resolution HST
imaging and high-resolution near-infrared and optical spectroscopy has
found SSC sizes ($\sim$~1--5 pc) and masses ($10^5$--$10^6$
M$_{\odot}$) comparable with those of globular clusters.  Estimates of
cluster ages have shown the SSCs to be much older than the crossing
time, indicating that they are gravitationally bound.  In addition,
radial profiles of the clusters are best fit with the King models
\citep[e.g.,][]{mengel02} derived for globular clusters
\citep{king62}.  These observations suggest that SSCs are young
globular clusters, forming in the present-day universe.

While SSCs have been identified in a number of galaxies, kinematic
masses have been measured for relatively few.  Among these are dwarf
galaxies NGC 1569 \citep{hoflipper96a,gilbert01} and NGC 1705
\citep{hoflipper96b}, and the NGC 4038/4039 merger \citep{mengel02}.
While much of this work has relied upon optical spectroscopy, there
are distinct advantages in the near-infrared.  SSCs are often situated
in dusty regions where extinction is high; the most heavily-reddened
clusters are not seen in optical images.  In addition, the spectra of
SSCs older than $\sim 7$ Myr are dominated by the light of cool
supergiant stars.  The near-IR spectra of these evolved stars feature
a multitude of atomic and molecular absorption lines which are
suitable for measurement of the stellar velocity dispersion within a
cluster via spectral cross-correlation.  Combined with high-resolution
images, this technique allows determination of cluster masses.

Kinematic masses for SSCs are relevant to the debate over the form of
the initial mass function (IMF) in starburst galaxies
\citep[e.g.,][]{eisenhauer01,gilmore01}.  Notably, \citet{rieke93} use
population synthesis models for the nuclear starburst in M82 to
conclude that the IMF is deficient in low-mass stars ($< 3
$M$_{\odot}$) relative to the IMF of the solar neighborhood.
\citet{sternberg98} finds that SSC NGC 1705-1 appears to be deficient
in low-mass stars, whereas a large fraction of the stellar mass in the
SSC NGC 1569A is contained in low-mass stars.  Measurement of the
kinematic mass of an SSC provides the only means of characterizing the
low-mass stellar population of the cluster.  While the light of these
stars is overwhelmed by the far more luminous supergiants, the
low-mass stars comprise a substantial fraction of the cluster's mass
for a normal IMF.  Quantifying the low-mass stellar content is
critical to constraining the IMF as well as assessing whether a
cluster can remain bound as a globular cluster.

M82 (NGC 3034) provides an interesting case.  As one of the nearest starburst
galaxies (3.6 Mpc) it has been extensively observed from X-rays to
radio wavelengths.  Its proximity results in an obvious resolution
advantage: star-forming regions can be studied on small spatial scales
($1'' = 17.5$ pc) and individual SSCs are resolved by HST.  A NICMOS
F160W mosaic (Figure \ref{sscmap}) of the nuclear starburst region
shows many luminous SSCs within $\sim 300$ pc of the nucleus.  The
nuclear starburst is thought to be active in the sense that the
typical age for the starburst clusters is $\sim 10^7$ years
\citep{satyapal97}.  Evolutionary synthesis models by
\citet{natascha98} suggest the nuclear starburst (i.e., the central
450 pc) consists of two distinct events with ages of about 5 Myr and
10 Myr.  \citet{o'connell95} imaged M82 in the $V$- and $I$-bands with
the high-resolution Planetary Camera aboard HST, identifying over 100
SSCs within a few hundred parsecs of the nucleus.  \citet{degrijs01}
imaged a region in the disk of M82, 1 kpc from the nuclear starburst,
with WFPC2 and NICMOS.  They identified 113 SSC candidates which they
believe were part of a starburst episode $\sim$ 600 Myr ago (a
``fossil starburst'').  \citet{smith01} derived the kinematic mass of
SSC `M82-F' using high-resolution optical (6010--8132\AA) spectroscopy
and WFPC2 (F439W) images.  M82-F (hereinafter MGG-F) lies $\sim$ 500
pc west of the nucleus of M82; Smith \& Gallagher estimate an age of
$60 \pm 20$ Myr, intermediate between the ongoing nuclear burst and
the fossil burst farther out in the disk.

In this paper, we investigate the kinematic mass of two clusters
(J0955502+694045, hereinafter MGG-11, and J0955505+694045, hereinafter
MGG-9) on the western edge of the active nuclear starburst.  These
particular clusters were chosen for this pilot program based on their
brightness ($m_H \sim 13$) and the ability to place both on a single
slit position.

\citet{freedman94} measured the distance to neighboring galaxy M81
using Cepheids, determining a value of $3.6 \pm 0.3$ Mpc.
Observations of HI \citep[e.g.,][]{appleton81} indicate a physical
link between M81 and M82.  As the projected separation between the
galaxies is only $\sim 40$ kpc, well within the uncertainty on the
distance, we adopt this value for M82.  The corresponding distance
modulus is $m - M = 27.8$.

\section{Observations}
\label{obs}

We observed super star clusters MGG-9 and MGG-11 with the 10-m Keck II
telescope on 2002 February 23, using the facility near-infrared
echelle spectrometer NIRSPEC \citep{mclean98}.  We obtained
high-resolution ($R \sim 22,000$), cross-dispersed spectra in the
wavelength range 1.51--1.75 $\mu$m using the NIRSPEC-5 order-sorting
filter.  The data fall in seven echelle orders, ranging from 44
through 50.  The clusters were imaged in four nods along the $0.''432$
$\times$ $24''$ slit, with successive nods separated by $5''$ (spatial
resolution of the spectra is $0.''5$).  Each integration was 600
seconds, providing a total time of 2400 seconds on the clusters.  The
spectra were dark subtracted, flat-fielded and corrected for cosmic
rays and bad pixels.  The curved echelle orders were then rectified
onto an orthogonal slit-position/wavelength grid based on a wavelength
solution from sky (OH) emission lines.  Each pixel in the grid has a
width of $\delta\lambda = 0.19$\AA.  We sky-subtracted by fitting
third-order polynomials to the 2D spectra column-by-column.  The
individual extracted spectra had average signal-to-noise ratios of
$\sim 47$ for MGG-9 and $\sim 42$ for MGG-11.

The NIRSPEC echelle turret is jostled whenever the cryogenic image
rotator is slewed.  When large slews, with amplitudes comparable to
the permitted range of motion, are executed, the echellegram shifts by
several pixels.  However, when the image rotator is making the small,
slow motions necessary to remove sidereal rotation, the wavelength
solution is stable to better than a few hundredths of a pixel.  This
is true even when an object is crossing the meridian and the sidereal
rotation is fastest.  Thus the velocity broadening reported here is
intrinsic to the source and not an instrumental effect.

The cluster spectra were extracted using Gaussian weighting functions
matched to the wavelength-integrated profiles.  To correct for
atmospheric absorption, we observed a B8V star, HD 74604.
Both the clusters and the calibration star were observed at an airmass
of $\sim 1.6$.  To account for photospheric absorption features
(particularly Brackett and helium lines) and continuum slope, the
calibration star spectrum was divided by a spline function fit.  The
resulting atmospheric absorption spectrum was then divided into the
cluster spectra.

We also observed several late-type stars of luminosity class I, II or
III with NIRSPEC for use as spectroscopic templates in
cross-correlation analysis (see Section \ref{spect}).  Table
\ref{calstars} summarizes these observations.  Template stars were
observed with the same slit, filter, and cross-disperser and echelle
grating positions as the clusters.  Reduction, rectification and
extraction steps were identical to those described above for the
clusters, with the exception that sky subtraction for the high
signal-to-noise stars was performed by pairwise subtraction of
successive nods along the slit.

For photometry and determination of structural parameters of super
star clusters in M82, we obtained images from the HST archive (HST
Proposal 7218, Marcia Rieke, PI).  The images were taken 1998 April 12
with the NIC2 camera on NICMOS and cover the nuclear region with four
adjacent, overlapping fields along the major axis.  We used the images
taken with the broadband F160W and F222M filters.  NIC2 has a plate
scale of 75 mas per pixel, and is therefore critically sampled at
1.75$\mu$m.  Each field was observed in a spiral dither of four
images, resulting in a field of view of $\sim 19.''5 \times 19.''5$
for each pointing center.  We relied upon the NICMOS data reduction
pipeline (calnicb) for combination of the dithered images.  The dither
pattern used a point-spacing of $0.''215$ (2.87 pixels).  At the
distance of M82, one pixel corresponds to 1.3 pc.  A mosaic of the
four fields covers an area of approximately $29'' \times 72''$.  Total
exposure time for each of the four fields was 192 seconds in F160W and
448 seconds in F222M (except the western-most field, which was 576
seconds).  We relied on the standard pipeline reduction procedures
(calnica) for dark-subtraction, flat-fielding and cosmic ray
rejection, and did not perform any additional calibration steps
ourselves.

We compared the astrometry of identifiably common clusters between the
NICMOS images and the positions listed in \citet{kronberg72}, and
found discrepancies of $\sim 2''$.  All astrometry quoted in this
paper has been adjusted to match Kronberg et al., and is
correspondingly accurate to $\sim 0.''6$.

\section{Cluster Velocity Dispersion}
\label{spect}

In the near-infrared, SSC spectra are dominated by the light of cool
evolved stars, the spectra of which contain a large number of
molecular and atomic absorption features \citep{gilbert01}.  We have
assembled high-resolution NIRSPEC spectra of a number of evolved
(luminosity class I, II and III) stars, ranging in spectral type from
G2 through M4.5.  Cross-correlation of these template spectra with the
spectrum of an individual super star cluster allows identification of
the dominant spectral type of stars in the SSC as well as
determination of the line-of-sight velocity dispersion.  For each
cluster in the present work, we have 28 measurements to
cross-correlate with the template star spectra: four nods in each of
seven echelle orders.  We have chosen not to sum the spectra for a
given echelle order to avoid the interpolation required in registering
the separate measurements to a consistent wavelength grid and the
resulting smearing of the spectral resolution.

Certain atmospheric OH emission lines were incompletely removed in the
sky subtraction process.  The residuals of these bright lines,
generally close doublets, must be removed prior to cross-correlation
to avoid introduction of systematic bias.  We have replaced the pixels
affected by residual sky emission with the median value of the five
pixels on either side of the contaminated range.  The fraction of
pixels replaced in a given echelle order typically amounts to a few
percent.  Tests on an OH emission-free echelle order indicate that the
cross-correlation result is unaffected within the stated
uncertainities, with the following exception.  At the systemic
velocity of M82, the bright telluric OH $4-2$ P2(3.5) emission line at
$1.6194 \mu$m overlies the CO ($v=3-6$) bandhead in echelle order 47
($1.609\mu$m $ < \lambda < 1.636\mu$m).  The CO bandhead feature is so
prominent in the spectrum that residual noise from sky subtraction
introduces significant systematic error for this echelle order.  The
results of the cross-correlation analysis for echelle order 47 are
therefore unreliable, and have not been used in this analysis.

The remaining 24 spectra for each cluster are cross-correlated with
the corresponding echelle orders of the template stars, and the
template star spectra are auto-correlated.  Our cross-correlation is
based on the method of \citet{tonry79}.  We fit a fifth-order
Chebychev polynomial to the spectrum to subtract the continuum, and
apodize with a cosine bell to remove mismatch at the edges.  The
spectrum is then transformed into Fourier space, and multiplied by
fourth-order Butterworth filters to remove low-frequency continuum
variations and high-frequency noise.  The filter frequencies are
selected to minimize any systematic bias introduced by the presence of
noise.  By the convolution theorem, multiplication (in Fourier space)
of the continuum-subtracted, apodized, filtered spectrum and its
complex conjugate, followed by inverse Fourier transformation, gives
the cross-correlation function.  Invoking the central limit theorem,
we fit the auto- and cross-correlation peaks with Gaussians to
determine their respective widths.  The excess width of the
cross-correlation relative to the auto-correlation of a template star
is a measure of the stellar velocity dispersion within the cluster.
Cross-correlation of a pair of template stars should reflect only
differences in rotation ($v \sin i$), macroturbulence, and noise.  The
advantage of using the auto-correlation is that it fully accounts for
the intrinsic width of absorption features when the correct template
spectral type is used.  Only the dispersion of photospheric
macroturbulence within that spectral type and luminosity class affects
our analysis, and no first-order correction to the measured velocity
dispersion is necessary.  The observed macroturbulence for K
supergiants is 11 km s$^{-1}$ with a peak-to-peak range of $\sim 2$ km
s$^{-1}$ \citep{gray87}.  Our analysis assumes that our template stars
are representative of their spectral type and luminosity class.

One potential difficulty is metallicity differences between Galactic
supergiants used as templates and the supergiants producing the
cluster light.  \citet[][and refs.]{mcleod93} cited evidence from
emission-line studies of various elements and concluded that the
present-day ISM in M82 has solar or slightly higher metallicity.
\citet{natascha01} determined that near-IR stellar absorption features
observed in the starburst core are consistent with the light from
solar-metallicity red supergiants (RSG).  Metallicity factors are
unlikely to significantly bias our measured velocity dispersions.

\begin{figure}
\centering \epsscale{1.0}
\plotone{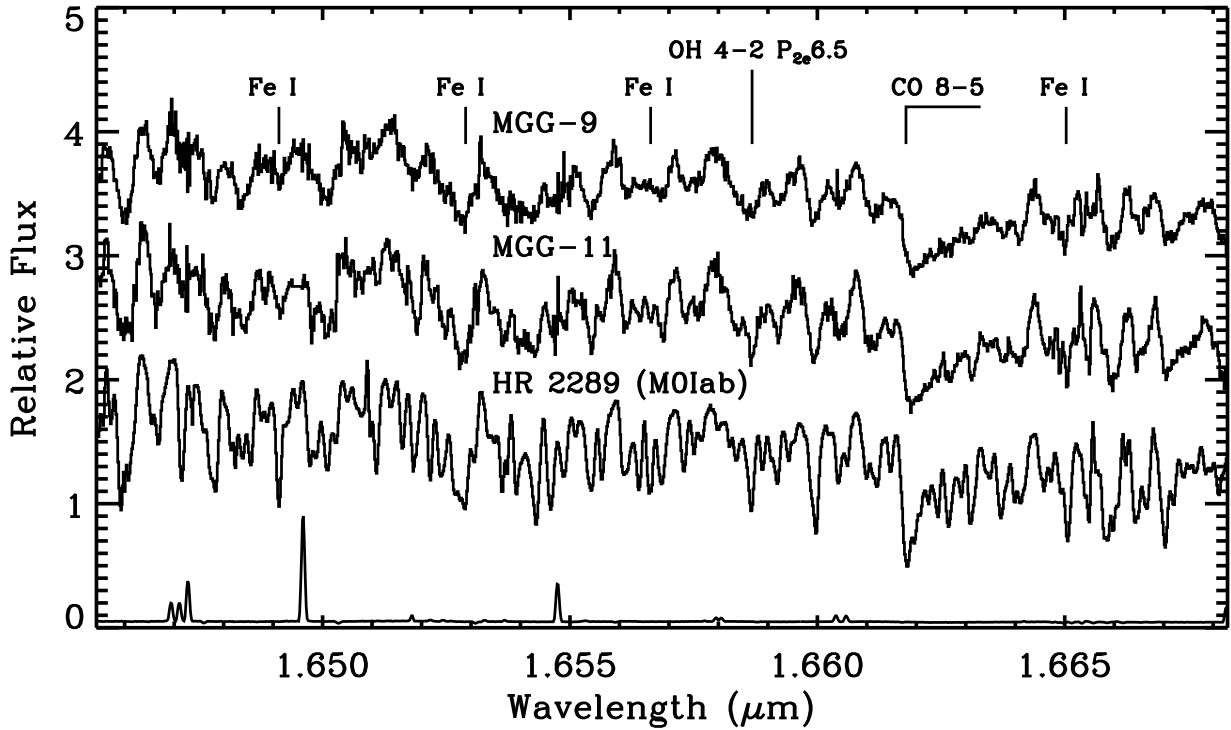}
\caption{Near-infrared rest-frame spectra of echelle order 46 for the
clusters and the M0Iab template star.  The spectra are identically
normalized, but offset vertically for clarity (for reference, zero
points are $-$0.9, 0.2 and 1.2, from bottom to top).  Many of the
features visible in the supergiant spectrum are evident in the SSC
spectra, although they appear ``washed out'' by the stellar velocity
dispersion of the cluster (e.g., the CO bandhead at $1.6618 \mu$m).
The sky emission spectrum is plotted at the bottom for reference.}
\label{ord46}
\end{figure}

\begin{figure}
\centering
\epsscale{1.0}
\plotone{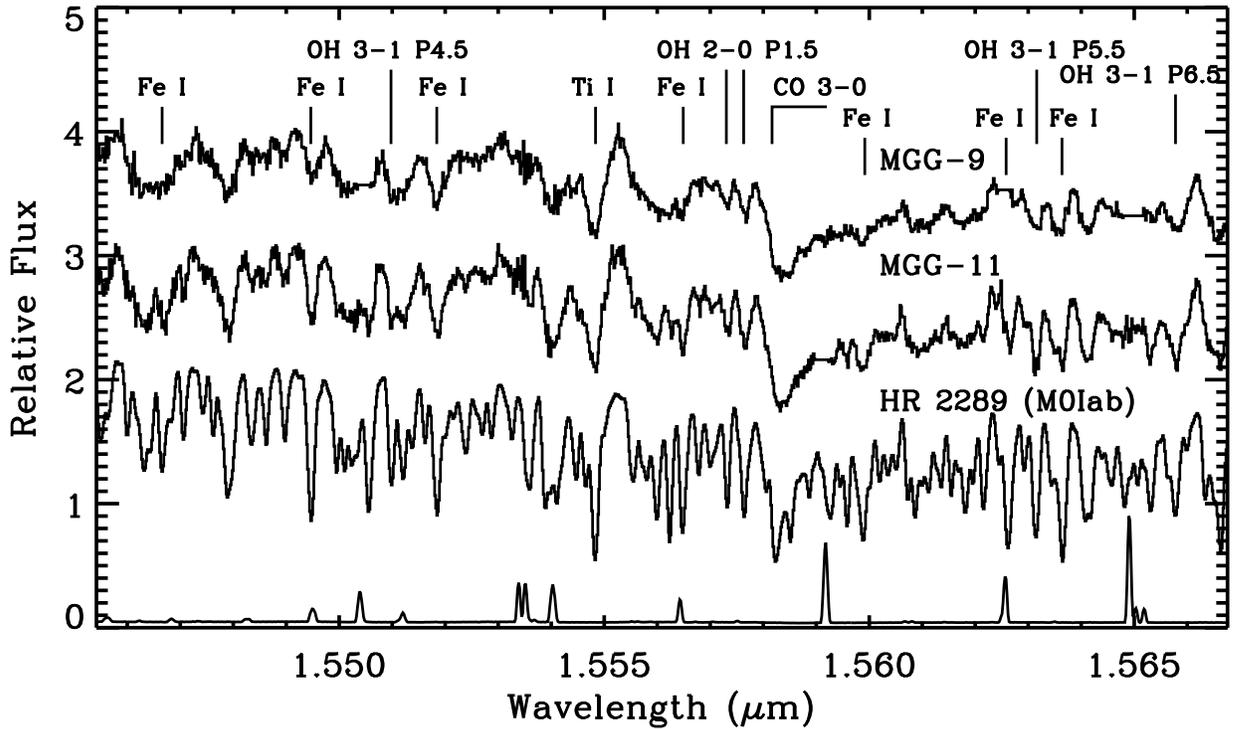}
\caption{Near-infrared rest-frame spectra of echelle order 49 for the
clusters and the M0Iab template star.  The spectra are presented in
the same manner as Figure \ref{ord46}.  This order is dominated by
strong metal absorption lines, including Ti~I at 1.5548 $\mu$m.  The
CO 3-0 bandhead at 1.5582 $\mu$m is also prominent.  The sky emission
spectrum is plotted at the bottom for reference.  Pixels replaced by
the local median value (Section \ref{spect}) may be identified by
comparison of the cluster and sky spectra.}
\label{ord49}
\end{figure}

\begin{figure}
\centering
\epsscale{0.8}
\plotone{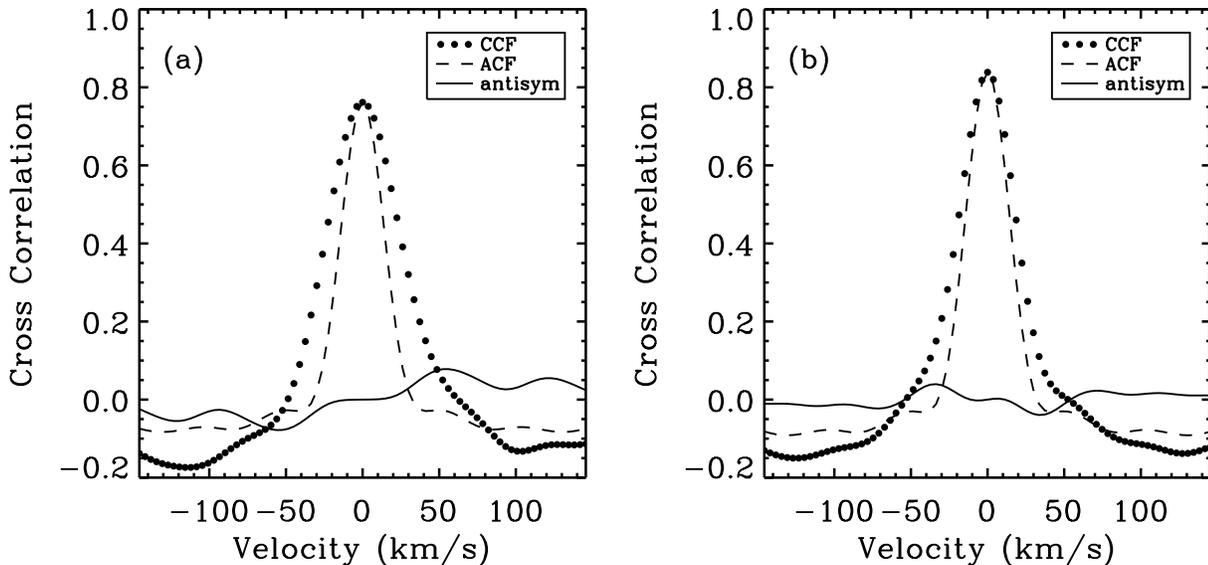}
\caption{Cross-correlation results for the clusters and the template
M0Iab star, averaged over all observed echelle orders (Section
\ref{spect}).  The peak of the template star auto-correlation function
(ACF) has been scaled to the peak of the CCF to emphasize the
difference in width.  The CCF is wider than the ACF due to the
convolution with the velocity distribution function of the cluster.
(a) CCF for MGG-9 and the template M0Iab star HR 2289.  The ACF peak
of the template is clearly narrower than the CCF peak.  The solid line
is the antisymmetric part of the CCF, showing a slight asymmetry.  (b)
CCF for MGG-11 and the same template M0Iab star, HR 2289.  MGG-11 has
a smaller velocity dispersion than MGG-9, as demonstrated by the
smaller difference in the widths of the ACF and CCF.  The CCF for
MGG-11 is highly symmetric.}
\label{avgccf}
\end{figure}

The results of the cross-correlation with the template stars are
presented in Table \ref{xcorr}.  The peak amplitude of the
cross-correlation function (CCF) measures the similarity of the
cluster to the template spectral type.  The CCF peak increases
steadily with spectral type from 0.3 for the G2I template to 0.8 for
the M0I and M4.5I templates.  We find that the spectra of clusters
MGG-9 and MGG-11 closely match the spectrum of the M0Iab (e.g.,
Figures \ref{ord46} and \ref {ord49}) and M4.5Iab stars.  The average
of the CCF over the 24 spectra for each cluster is shown in
Figure~\ref{avgccf}.  Unfortunately, our spectral atlas does not
presently include any template supergiant spectra between M0 and M4.5.
Intermediate spectral types, e.g. M2, should match the clusters at
least as well as M0 and M4.5.

Asymmetries in the CCF indicate kinematic substructure, such as the
presence of an additional cluster.  NGC 1569A, for example, displays
an asymmetric CCF which is the result of a kinematically distinct
cluster which is projected onto the same slit position
\citep{gilbert02iau, hoflipper96a}.  The CCFs for these M82 SSCs are
quite symmetrical, and are well fit by Gaussians.  The clusters show
evidence of only a single kinematic component.

Based on an average of the results from the best-match templates, we
measure velocity dispersions of $\sigma_r = 15.9 \pm 0.8$ km s$^{-1}$
for MGG-9 and $\sigma_r = 11.4 \pm 0.8$ km s$^{-1}$ for MGG-11.  (The
uncertainties quoted here are the standard errors --- the standard
deviation of the independent measurements divided by $\sqrt{N}$.)

Cross-correlation with a mismatched template spectrum can introduce
systematic bias to the velocity dispersion determination.  We note a
roughly linear trend between the measured velocity dispersion and the
peak value of the CCF for a given template.  In the case of MGG-11,
extrapolation of the trend to a hypothetical ``perfect'' correlation
of 1.0 gives a value of $\sigma_r$ which is consistent with the value
quoted above.  In the case of MGG-9, the strong decreasing trend of
$\sigma_r$ with the CCF peak value suggests that the quoted value may
be slightly inflated due to systematic bias from template mismatch.
Comparison to a correlation of 1.0, however, is not the correct
metric; the velocity dispersion of the cluster stars reduces the CCF
peak even for a perfect spectral match.  As a test, we convolved the
M0Iab spectrum with a Gaussian velocity distribution, and then
cross-correlated the result with the template M0Iab spectrum.  The CCF
peaked at 0.90 for a velocity dispersion of 16.2 km s$^{-1}$;
increasing the velocity dispersion decreases the CCF peak (note the
higher CCF peak values for MGG-11, which has a lower velocity
dispersion than MGG-9).  We adopt the quoted $\sigma_r$ values with
the caveat that there may be a small amount ($\sim$ 1 km s$^{-1}$) of
systematic bias present due to template mismatch.

Cross-correlation analysis also measures the heliocentric radial
velocities of the clusters.  We find that MGG-9 has $v_r = 113.8 \pm
0.4$ km s$^{-1}$ and MGG-11 has $v_r = 92.5 \pm 0.5$ km s$^{-1}$.
These values are based on cross-correlation with HD 43335, which has a
radial velocity of 38.9 km s$^{-1}$ (per
SIMBAD\footnote{http://simbad.u-strasbg.fr/}).  The spectrum of HD
43335 is was taken immediately before the cluster spectra, and we find
no grating shifts which would systematically bias the radial
velocities.

Millimeter observations of CO emission by \citet{shenlo95} provide
context for the radial velocities.  The clusters lie in a ``saddle''
feature in the velocity-integrated CO emission map, roughly $10''$
west of the kinematic center.  Plotted on Shen \& Lo's major axis
$p$--$v$ cut, the clusters lie on the western shoulder of a prominent
second spatial component, separate from the primary velocity gradient.
The radial velocities of the clusters measured in the near-IR are
consistent with the velocity of CO at their locations, suggesting a
potential kinematic link between the clusters and their natal
molecular gas.  Deprojecting the galaxy and applying the Shen \& Lo CO
velocity gradient to a simple cylindrical solid-body rotation model
gives an orbital period of $\sim 17$ Myr for the clusters.  They have
not had sufficient time to complete an orbit of the galaxy, and may
yet be near their birthplaces.  Future high-resolution studies will
shed more light on the kinematic connection between dense molecular
gas and young super star clusters in the starburst.

\section{Cluster Integrated Luminosity}
\label{hst}

\begin{figure}
\centering
\epsscale{1.0}
\plotone{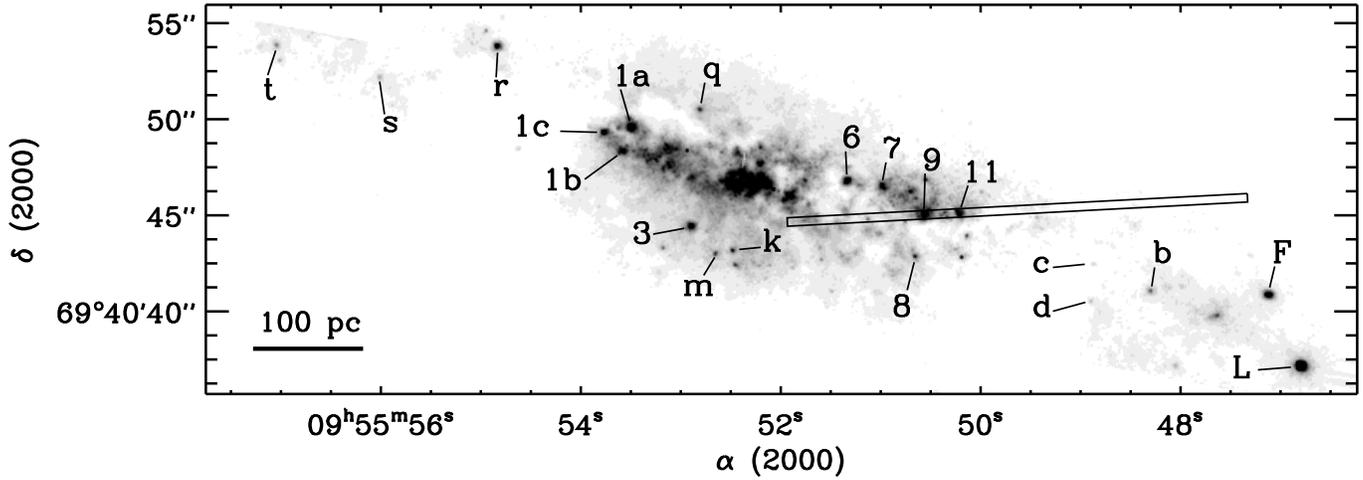}
\caption{Mosaic of NICMOS F160W images of the nucleus of M82.
Clusters included in Table \ref{sscdata} are labelled with the
secondary identifier for reference.  One of the four nod positions of
the $24'' \times 0.''432$ slit is shown.  At the distance of M82, $1''
= 17.5$ pc; the scale bar at lower left is 100 pc long.  }
\label{sscmap}
\end{figure}

Figure \ref{sscmap} shows the NICMOS F160W mosaic of the nuclear
starburst in M82, with the clusters treated in this paper labelled
with our secondary designations.  The region is notoriously complex,
showing structure on virtually any spatial scale resolvable.  In the
NICMOS images, many of the candidate super star clusters appear
against a mottled background of relatively strong diffuse emission.
In order to perform photometry and measure size parameters, we must
isolate the cluster light from the background.  We placed a
0.$''$15-wide annulus at an inner radius of 0.$''$5 (9 pc) from the
centroid of each cluster, and fit a minimum-least-squares plane to the
points within the annulus.  Subtraction of this plane removes the
background emission, allowing us to fit the cluster light.

\begin{figure}
\centering
\epsscale{1.0}
\plotone{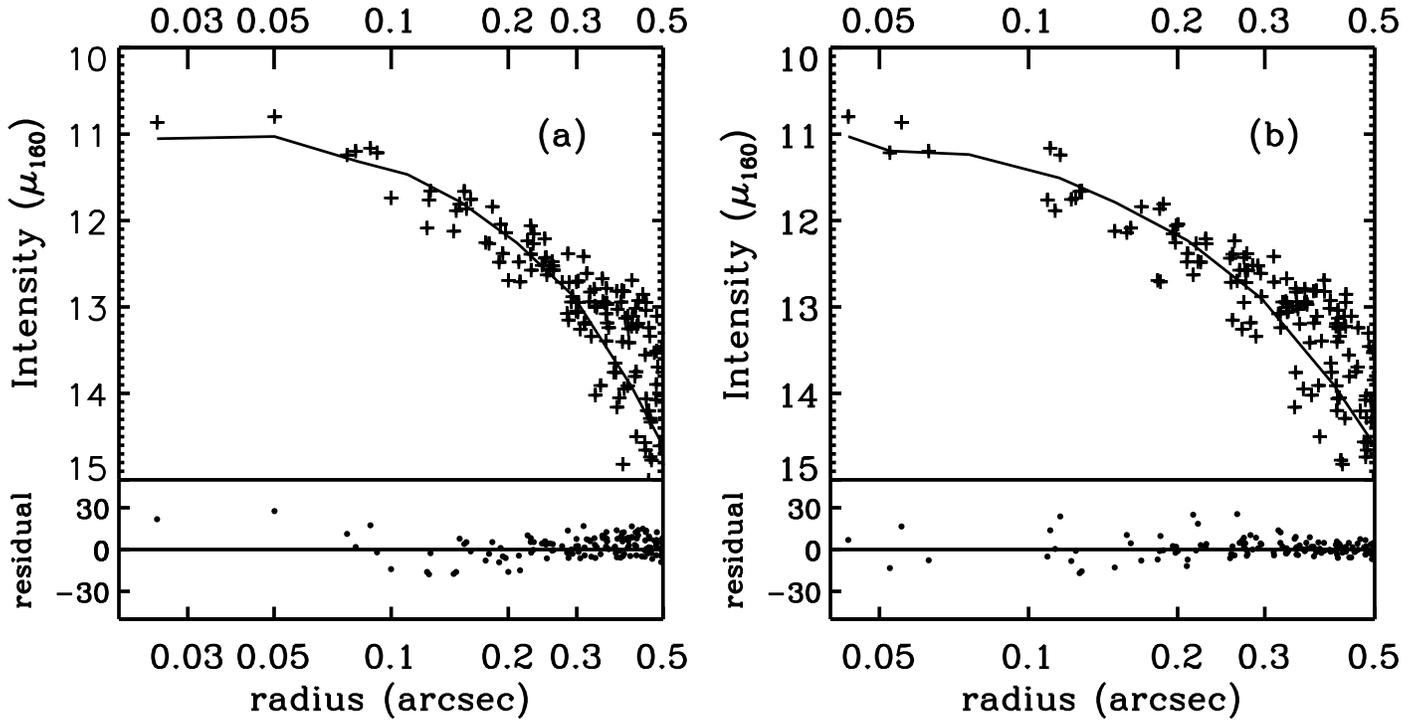}
\caption{ (a) Radial profile plot for MGG-9.  The sky-subtracted
NICMOS data ($+$) were fitted in two dimensions, but are collapsed
azimuthally for this plot.  The ordinate is plotted in [F160W]
magnitudes per square arcsec ($\mu_{160}$).  The solid line shows the
azimuthal average of the fit, which is a King model convolved with the
NIC2 PSF.  The lower plot shows the two-dimensional, point-to-point
residuals (data minus fit) in NIC2 counts/sec. (b) Radial profile plot
for MGG-11. }
\label{radial}
\end{figure}

Calculation of the virial mass of a cluster (Section \ref{dynmass})
requires determination of the half-light radius.  Special care is
required, in that many of the clusters are only marginally resolved.
The effect of the instrumental point spread function must therefore be
taken into account.  For each cluster, we generated a model NIC2 PSF
using the Tiny Tim program \citep{Krist95}.  The PSF models were
matched to the centroid location of each cluster on the detector,
oversampled $2\times$ and rebinned to match the pixel size of NIC2.
To model the observed cluster images, we tested various functional
forms of the radial brightness profile.  We found good fits to the
data for the empirical King model \citep{king62} of the form:

\[
f(r) = k \left\{
         \frac{1}{\left[1+(r/r_c)^2\right]^{\frac{1}{2} } }-
         \frac{1}{\left[1+(r_t/r_c)^2\right]^{\frac{1}{2} } }
         \right\}^2
\]

\noindent where $0 \le r \le r_t$.  The King model has three free
parameters: the {\it core radius} $r_c$, the {\it central
concentration} $r_t/r_c$ (where $r_t$ is the tidal radius), and a
scaling factor $k$.  The King model is convolved with the PSF model
and compared to the data.  The best-fit model is determined using a
Levenberg-Marquardt minimization, finding the least squares fit to the
data through iterative search of parameter space.  The resulting
models give very good fits, as illustrated in Figure \ref{radial}.
Once the King model parameters have been fit for a given cluster, the
projected half-light radius ($r_{hp}$) is the numerical solution of

\[
\frac{\int_0^{r_{hp}} f(r)2\pi r dr}{\int_0^{r_t} f(r)2\pi r dr} =
\frac{1}{2}
\]

Likewise, integration of the fitted King model over the interval $0
\le r \le r_t$, in conjunction with the NIC2 photometric keyword
PHOTFNU, provides the total apparent magnitude of the cluster.  Table
\ref{sscdata} lists the astrometry, photometry and derived half-light
radii for the 20 NICMOS-resolved clusters labelled in Figure
\ref{sscmap}.  In particular, we find projected half-light radii
($r_{hp}$) of $146 \pm 16$ mas for MGG-9 and $66 \pm 7$ mas for
MGG-11.  At the adopted distance of $3.6 \pm 0.$ Mpc, the projected
half-light radii are $2.6 \pm 0.4$ pc for MGG-9 and $1.2 \pm 0.17$ pc
for MGG-11.

To determine the uncertainties associated with the photometry and
half-light radii, we performed a Monte Carlo simulation wherein we
added fake star clusters to the NICMOS images and used our fitting
algorithm to recover them.  We tested fake clusters with a range of
luminosities and King model parameters, placed against a
representative range of background conditions.  While the algorithm
could not reliably recover the values of $r_c$ and central
concentration ($r_c/r_t$), it must be noted that these parameters are
highly covariant; a particular halflight radius may be represented by
a family of $r_c$ and $r_t$ values.  The fitting algorithm identified
the projected halflight radii to within an uncertainty of 12\%,
independent of the input values of $r_c$ and $r_t$.  Photometric
uncertainities scaled with cluster luminosity, ranging from 0.03 to
0.4 mag in low background conditions.  In regions of high and/or
variable background emission, the algorithm produced larger scatter
around the correct luminosity.  The results of this analysis are
reflected in the photometric uncertainties in Table \ref{sscdata}.

We have chosen to use the F160W images for determination of the
half-light radii rather than the F222M images for two reasons: (1) we
are measuring the velocity dispersion of our test particle stars at
1.51 -- 1.75 $\mu$m, and it is consistent to determine structural
parameters of the test particles by using the same light; and (2) warm
circumstellar dust may result in excess continuum emission longward of
2 $\mu$m, complicating cross-correlation analysis.

MGG-F has been closely examined by other authors, and provides a
comparison for our measurements.  \citet{o'connell95} imaged MGG-F in
the $V$-band (F555W) with WFPC.  They deconvolve the images using a
Tiny Tim PSF model, and find a mean FWHM of 160 mas (2.8 pc).  The
authors note that the deconvolution is ``not highly reliable'' because
the cluster falls near the edge of the detector.  \citet{smith01}
performed circular aperture photometry on MGG-F in F439W images from
WFPC2 and find a projected half-light radius of $160 \pm 20$ mas ($2.8
\pm 0.3$ pc).  This value does not reflect deconvolution, but rather a
``simple approach'' of subtracting 150 mas from the observed value in
quadrature to correct for the PSF.  By comparison, our profile-fitting
method finds a smaller projected half-light radius of $89 \pm 11$ mas
($1.6 \pm 0.2$ pc) in both F160W and F222M.  \citet{sternberg98} notes
that if the massive red supergiants are spatially segregated from
intermediate mass stars, the SSCs will appear more compact in the
near-IR than in the visible.  The smaller size of MGG-F as measured in
the NICMOS images may thus be evidence of mass segregation.

\section{Discussion}
\label{analy}

The combination of spectroscopic and photometric data enables us to
examine the stellar composition of the clusters.  In particular, we
can use measured light-to-mass ratios to draw inferences on the form
of the cluster IMFs.  Using the stellar velocity dispersions and
half-light radii determined above, we find the cluster masses in \S
\ref{dynmass}.  Extinction to the clusters, which plays a critical
role in subsequent analysis, is estimated in \S \ref{extinct}, and
cluster ages are estimated in \S \ref{ages}.  The light-to-mass ratios
and IMF implications are discussed in \S \ref{m2l}, leading to
discussion of the fate of the clusters in \S \ref{fate}.

\subsection{Kinematic Masses}
\label{dynmass}

The mass of a star cluster may be determined by application of the
virial theorem, which dictates that for a cluster in equilibrium,
kinetic and potential energy are related by $V^2 = \eta GM/R$.
\citet{spitzer69} shows that when expressed in terms of the radius
containing half of the mass of the cluster ($r_h$), the constant of
proportionality is well approximated by the value $\eta = 0.4$ for
most mass distribution models.  Assuming that the light profile traces
the mass distribution, the half-light radius is used as a proxy for
the half-mass radius.  For an isotropic velocity dispersion, the
half-light radius is related to the observed half-light radius in
projection ($r_{hp}$) by $r_{hp} \sim \frac{3}{4}r_h$
\citep{spitzer87}, and the three-dimensional rms velocity $V$ is
related to the measured one-dimensional line-of-sight velocity
dispersion $\sigma_r$ by $V^2 = 3\sigma_r^2$.  Armed with
determinations of the velocity dispersion and projected half-light
radius, we are thus able to estimate the mass of a virialized super
star cluster:

$$
M = 10\,\frac{r_{hp} \sigma_r^2}{G}
$$

Under these assumptions, we find a mass of $1.5 \pm 0.3 \times 10^6$
M$_{\odot}$ for MGG-9.  This corresponds to an average mass density of
$1.1 \pm 0.3 \times 10^4$ M$_{\odot}$ pc$^{-3}$ within the half-mass
radius.  MGG-11 is less massive at $3.5 \pm 0.7 \times 10^5$
M$_{\odot}$, with a higher average mass density within $r_{hp}$ of
$2.7 \pm 0.9 \times 10^4$ M$_{\odot}$ pc$^{-3}$.

As noted in Section \ref{hst}, MGG-F shows evidence of mass
segregation.  While extensive dynamical mass segregation in MGG-9 and
MGG-11 is unlikely (Section \ref{ages}), it is possible that the most
massive stars formed nearer the cluster cores.  In this case, the red
supergiant velocity dispersions we measure in the near-IR would be
smaller than the cluster mean and the masses we derive would represent
lower limits.

\subsection{Extinction}
\label{extinct}

\begin{figure}
\centering \epsscale{0.9} 
\plotone{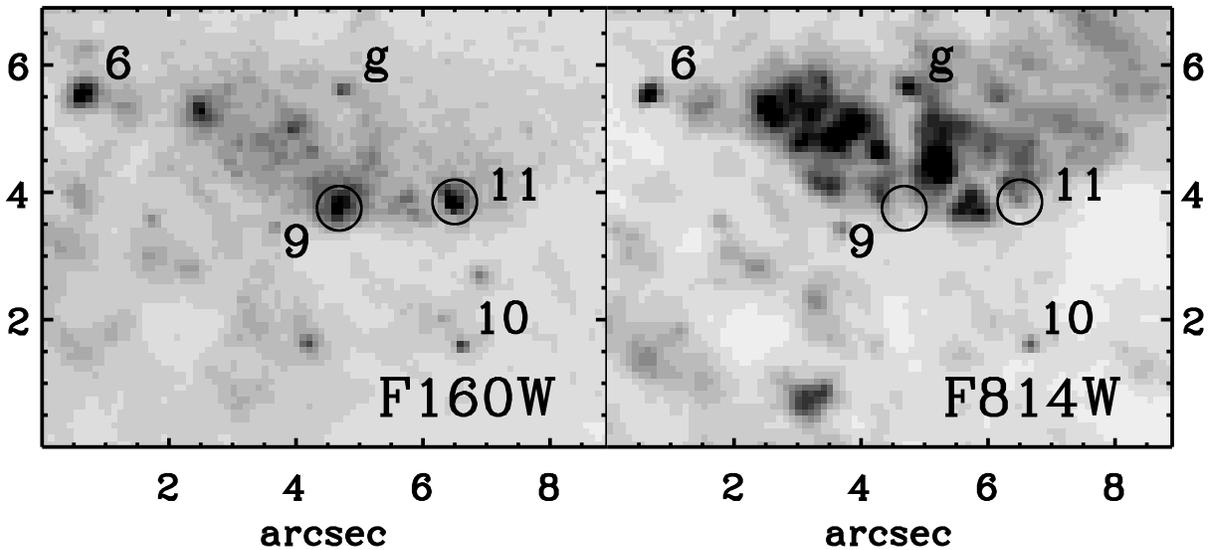}
\caption{A comparison of the NICMOS F160W image with an archival WFPC2
F814W image of the region surrounding clusters MGG-9 and MGG-11.
Circles mark the position of these clusters in each image; while there
is faint emission evident at the position of MGG-11 in the F814W
image, MGG-9 is completely invisible due to heavy extinction.  The
NICMOS image has been resampled to match the 0.1 arcsec pixels of
WFPC2.  Certain clusters have been labelled as reference points
between the images; all axes are labelled in arcsec.}
\label{hvsi}
\end{figure}

As noted in Section \ref{hst}, near-IR images of the nuclear starburst
show substantial diffuse emission which hints at the importance of
interstellar extinction (see also Figure \ref{hvsi}).  Theoretical
studies \citep[e.g.,][]{hills80} indicate that the star formation
efficiency is high ($> 50\%$) in massive star clusters.  Considering
the additional gas-dispersing effects of supernovae and stellar winds
from young, high-mass stars, it is unlikely that much gas from the
natal cloud remains in the clusters after a few million years.  We
have thus assumed a foreground screen model for our extinction
estimates.

The spectra of both MGG-9 and MGG-11 are dominated by the light of
early M-type supergiants.  Based on the intrinsic color of an M2I star
(representing the midpoint of the M0I -- M4.5I range), we can estimate
extinction along the line of sight to the clusters.  We calculated
synthetic magnitudes for stars in the \citet{pickles98} spectral
library for NICMOS filters.  The Pickles library includes an M2I star,
which has a synthetic [F160W] $-$ [F222M] color of 0.52.  We adopt
this as the intrinsic color of the clusters, and allow for a scatter
of $\pm 0.1$ when calculating uncertainties.

\begin{figure}
\centering \epsscale{0.6} 
\plotone{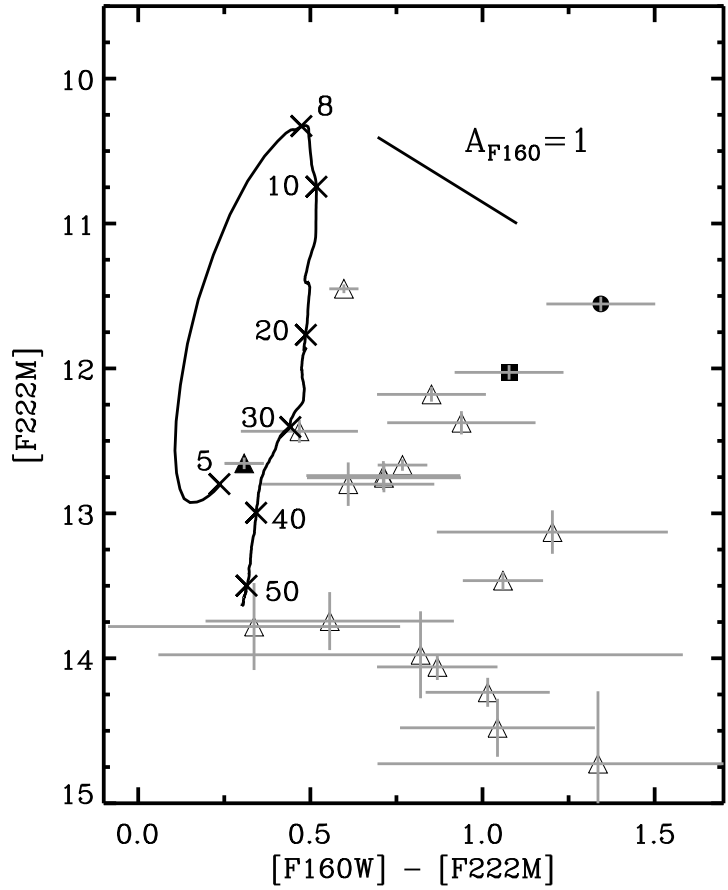}
\caption{Color-magnitude diagram of the SSCs included in Table
\ref{xcorr}.  MGG-9 (filled circle) is the reddest cluster in the
sample, and MGG-11 (filled square) is also more red than the average;
MGG-F (filled triangle) is the bluest cluster.  The Starburst99
evolutionary track for a $10^6$ M$_{\odot}$ cluster with a Kroupa IMF
is displayed with cluster ages labelled in millions of years.  Note
that the magnitude of a cluster is dependent upon its mass; the age of
a cluster cannot be directly determined from the plotted track without
knowledge of its mass. }
\label{cmd}
\end{figure}

Evolutionary tracks from population synthesis models predict a
similar color ([F160W] $-$ [F222M] $\sim 0.5$) for the clusters, as
seen in the color-magnitude diagram (CMD) for the nuclear starburst
(Figure \ref{cmd}).  The Starburst99 \citep{leitherer99} models used
generate $H$ and $K$ magnitudes on the Johnson system.  We used the
method of \citet{marleau00} to determine the color transformation
between the Starburst99 output and the NICMOS filters:

$$
(\mbox{[F160W]}-\mbox{[F222M]}) = 1.46(H - K)_J + 0.042
$$

Using the data in Table \ref{sscdata}, the [F160W] $-$ [F222M] colors
of MGG-9 and MGG-11 are $1.35 \pm 0.16$ and $1.07 \pm 0.16$,
respectively.  Subtracting the adopted intrinsic color of the
clusters, we find color excesses of E([F160W] $-$ [F222M]) $= 0.83 \pm
0.19$ for MGG-9 and $0.55 \pm 0.19$ for MGG-11.  Applying the
extinction law of \citet{cardelli89}, we estimate the extinction in
the F160W bandpass as $A_{F160W} = 2.47
E(\mbox{[F160W]}-\mbox{[F222M]})$.  Thus we find $A_{F160W} = 2.1 \pm
0.5$ for MGG-9 and $1.4 \pm 0.5$ for MGG-11.

As noted by Cardelli et al., there exists a single extinction law for
$\lambda \ge 0.90 \mu$m.  As such, the estimated extinction
$A_{F160W}$ is independent of assumptions regarding the value of
$R_V$.  For reference, if $R_V = 3.1$, $A_{F160W} = 0.19A_V$ and if
$R_V = 5$, $A_{F160W} = 0.22A_V$.  The clusters plotted in Figure
\ref{cmd} display differential extinction, from zero to as high as
perhaps 12 magnitudes in $A_V$.  This is not surprising given the
range of background intensities of the diffuse emission seen in the
NICMOS images (Figure \ref{sscmap}).  Extinction plays a critical role
in determining the light-to-mass ratios of the clusters (Section
\ref{m2l}).  Observers should be wary when attempting to trace star
formation with UBV photometry and optical spectra in dusty starburst
and interacting galaxies.

\subsection{Ages}
\label{ages}

The ages of the SSCs may be constrained by several inferences drawn
from their spectra.  The absence of nebular emission lines indicates
that there are no O stars remaining in the clusters.  This sets a
lower age limit of about 6--7 Myr, the main sequence lifetime of O
stars.  As noted above, the spectra are dominated by the light of red
supergiant stars as evidenced by the strong photospheric CO
absorption.  This is broadly consistent with somewhat older starburst
populations (ages $\sim 10^7$ yrs).  

Cool, intrinsically luminous asymptotic giant branch (AGB) stars
contribute substantially to the near-IR light of clusters between ages
0.3 to 1.5 Gyr, dominating the near-IR light at $\sim 600$ Myr
\citep{mouhcine02}.  The thermally-pulsing AGB (TP-AGB) phase begins
to significantly affect the near-IR colors of a population at an age
of about 100 Myr.  The TP-AGB phase is not modelled by the Geneva
evolutionary tracks used in Starburst99, compromising the
applicability of the population synthesis models beyond this point.
The deep CO bands of TP-AGB stars could be mistaken for features of
RSG stars in a younger population in the absence of other age
indications \citep{gilbert02th}.  Unfortunately, the best test for the
predominance of AGB stars, the H$_2$O/C$_2$ index \citep{lancon99},
falls just outside our spectral range at 1.77$\mu$m.  However,
\citet{natascha98} notes that none of the features expected of AGB
stars are present in her near-IR spectra of the nuclear region of M82,
setting an upper age limit of $\sim 10^8$ yrs.

We can estimate the age range more precisely through population
synthesis modelling.  \citet{gilbert02th} used Starburst99 models to
predict the flux-weighted average spectral type as a function of
cluster age.  Applying this technique at $H$-band, we find that the
light of a cluster population is dominated by M0 stars, as indicated
by our cross-correlation analysis for these SSCs, at two different
ages in a Salpeter IMF model: 5--7 Myr and 12--14 Myr.  The former can
be ruled out in this case, as there are no O stars present.  During
the intervening period (7--12 Myr), the dominant spectral type in
$H$-band is later than M0, but is at no time later than M3.  The
latest template star in our high-resolution NIRSPEC spectral atlas is
type M4.5; this spectral type matches the cluster spectra as well as
the M0 template.  This suggests that the best match is likely M2 or
M3, spectral types as yet missing from our template atlas.  {\it We
therefore estimate ages of 7--12 Myr.}  This is consistent with the
10--12 Myr age of the ``older'' burst population found by
\citet{natascha98} for the nuclear starburst region.

Figure \ref{cmd} shows the CMD for the nuclear clusters.  For context,
we have overplotted an evolutionary track based upon a model $10^6$
M$_{\odot}$ cluster with a Kroupa IMF (Section \ref{m2l}) for stellar
masses of 0.1 to 100 M$_{\odot}$.  The clusters fall in a region which
is consistent with ages of 8 to 60 Myr, and exhibit differential
extinction (Section \ref{extinct}).  Ages cannot be read directly from
Figure \ref{cmd} because of the dependence of F222M luminosity on
total mass.  Dereddening MGG-9 until it meets the evolutionary track
for a cluster with its derived mass of $1.5 \times 10^6$ M$_{\odot}$,
we find a most likely age of approximately 10 Myr.  When dereddened,
MGG-11 is too bright for the Kroupa IMF evolutionary track
corresponding to its derived mass at any age.  Using the evolutionary
track of a Salpeter IMF truncated at a minimum mass of 1 M$_{\odot}$,
we find a most likely age for MGG-11 of approximately 9 Myr.  This is
discussed further in Section \ref{m2l}, along with predictions of the
time evolution of light-to-mass ratios.  It should be noted, however,
that these age estimates are based on an assumed IMF, and cannot
therefore be used to infer information on the stellar content.  For
this purpose, we will use the IMF independent age estimates of 7-12
Myr.

Measurement of the half-light radii and velocity dispersions provides
an estimate of the dynamical (crossing) times for the clusters.  Both
MGG-9 and MGG-11 have $t_{cr} \sim 10^5$ years.  The clusters are clearly
hundreds of crossing times old and thus gravitationally bound.
Calculating the characteristic relaxation time $t_r$ using the
prescription of \citet{meylan87}, we find $t_r > 10^8$ years for these
SSCs.  At the present young ages of the clusters, we do not expect
significant, cluster-wide dynamical mass segregation.  (See Section
\ref{m2l} for further discussion on mass segregation.)

\citet{smith01} estimated the age of MGG-F as $60 \pm 20$ Myr.  Our
population synthesis models indicate a younger age of $\sim 40$ Myr.
In the age range of 25--40 Myr, the cluster light in $H$-band is
dominated by the light of G-type stars.  This implies
E([F160W]--[F222M]) $\sim 0.1$ for MGG-F and thus $A_{F160W} \sim .2$,
in close agreement with the estimated $E(B-V) = 0.9 \pm 0.1$ of Smith
\& Gallagher.

\subsection{Light-to-Mass Ratios and the IMF}
\label{m2l}
The near-infrared light of SSCs older than $\sim 7$ Myr is dominated
by high-luminosity, massive, evolved stars.  Lower-mass stars
contribute very little to the spectrum of a young SSC, even if they
comprise a majority of the cluster's mass.  Measurement of the
cluster's kinematic mass enables detection of the presence of low-mass
stars {\it independent of any assumptions regarding the form of the
IMF}.  Quantification of the low-mass stellar content in turn provides
insight as to whether the SSC can remain gravitationally bound over a
globular cluster lifetime.

From our photometry, we determine the total luminosity of the clusters
in the F160W filter: $L_{F160} = 4\pi d^2 F_{\nu}\,\Delta\nu$ (where
$\Delta\nu = 1.37 \times 10^{13}$ Hz for the filter bandwidth of
$\Delta\lambda = 1177.3$\AA).  Before adjustment for interstellar
extinction, we find $L_{F160}/M$ of $0.28 \pm 0.06$ (in solar units,
L$_{\odot}$/M$_{\odot}$) for MGG-9 and $1.0 \pm 0.2$ for MGG-11.
Dereddening by the estimated extinctions (Section \ref{extinct}) gives
$L_{F160}/M$ of $1.8^{+1.1}_{-0.8}$ for MGG-9 and $3.5^{+2.1}_{-1.5}$
for MGG-11.

\begin{figure}
\centering \epsscale{0.7} 
\plotone{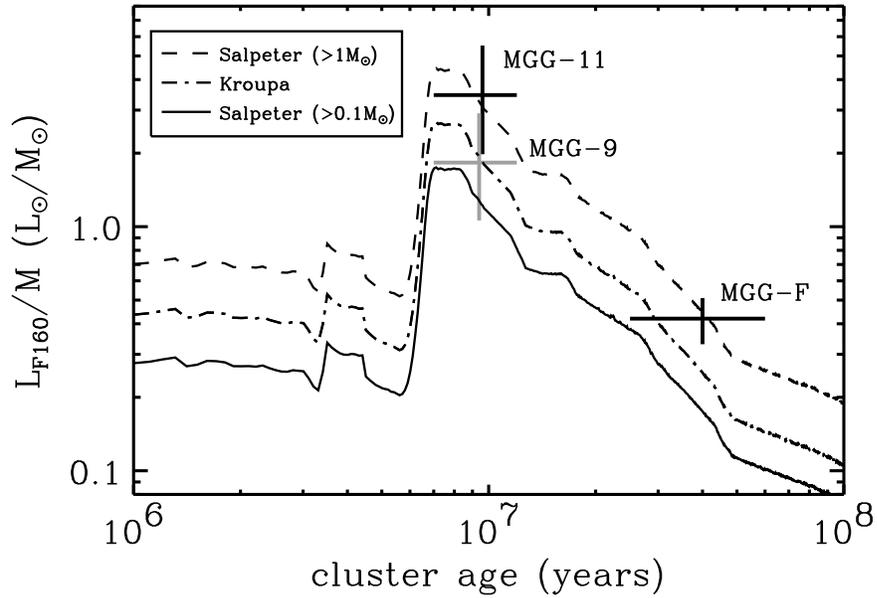}
\caption{Evolution of the luminosity-to-mass ratio as a function of
age based on Starburst99 modelling.  Three IMF models are shown for a
fiducial cluster mass of $10^6$ M$_{\odot}$: Salpeter single-power-law
IMFs with minimum stellar masses of 0.1 or 1.0 M$_{\odot}$, and a
Kroupa IMF (see Section \ref{m2l}).  MGG-9 appears consistent with a
Kroupa IMF, whereas MGG-11 and MGG-F may be deficient in low-mass
stars.  The vertical error bars for MGG-9 and MGG-11 demonstrate the
uncertainty on the $L/M$ ratio; their horizontal location does not
reflect any age information within the range of 7--12 Myr.  The MGG-F
data are based on the mass estimate of \citet{smith01}.  }
\label{starburst99}
\end{figure}

Figure \ref{starburst99} shows predicted $L_{F160}/M$ ratios as a
function of cluster age, based on population synthesis models.  We
have plotted models for three different IMF forms: a
\citet{salpeter55} power-law IMF with stellar masses of 0.1--100
M$_{\odot}$, a \citet{kroupa01} IMF, and a Salpeter power law
truncated with a minimum stellar mass of 1M$_{\odot}$.

The stellar IMF is more complex than a single power law IMF extending
down to the lowest stellar masses \citep[see review in][]{meyer00}.
In general, the IMF is represented as a broken power law, with one
slope for the distribution of high-mass stars and a lower slope or
slopes for the low-mass ($< 1 $M$_{\odot}$) stars.  The overall effect
of such a distribution is to reduce the relative proportion of
low-mass stars with respect to a single-slope power law distribution.
The Kroupa IMF is respresentative: $\xi(M) \propto M^{-\alpha_i}$
where

\begin{eqnarray*}
\alpha_1 = 1.3, \quad 0.1 \le M/\mbox{M}_{\odot} < 0.5 \\
\alpha_2 = 2.3, \quad 0.5 \le M/\mbox{M}_{\odot} \le 100 \\
\end{eqnarray*}

We find that the $L_{F160}/M$ ratio of MGG-9 is consistent with a
solar-metallicity Kroupa IMF for the cluster's age and dereddened
luminosity.  It does not appear that there is any dimunition in the
relative proportion of low-mass stars in this cluster.  MGG-11,
however, has a dereddened luminosity which is too bright for a Kroupa
IMF at any age.  MGG-11 appears to be deficient in low-mass stars.
This suggests either a non-standard IMF or mass segregation within the
cluster (see below).  For comparison, Figure \ref{starburst99} shows
the $L_{F160}/M$ ratio of a single-power law IMF (with $\alpha =
2.35$, as in a Salpeter IMF) which is truncated at a minimum mass of
1M$_{\odot}$.  Such a distribution of stellar masses would lead to a
$L_{F160}/M$ ratio consistent with our observations for MGG-11.

We also consider MGG-F.  Using the mass estimate of \citet{smith01} in
combination with our photometry, we find a dereddened $L_{F160}/M =
0.4 \pm 0.1$.  This light-to-mass ratio is likely too high for the
cluster's age and a standard Kroupa IMF, again suggesting a deficiency
of low-mass stars.  Smith \& Gallagher conclude that the lower mass
limit is 2--3M$_{\odot}$ for a power-law IMF of slope $\alpha = 2.3$;
our lower estimate of the age suggests that the lower mass cutoff need
only occur around 1M$_{\odot}$.  

As noted in Section~\ref{hst}, there is evidence of mass segregation
in MGG-F.  The relaxation timescale is too great for significant
dynamical mass segregation to have occurred throughout clusters MGG-9
and MGG-11 (ages $\sim 10^7$ yr).  Numerical simulations by
\citet{bonnell98} indicate that although clusters younger than their
relaxation timescale are not fully mass-segregated, they are likely to
show some measure of mass segregation at young ages.  Specifically,
mass segregation of the most massive stars is known to occur more
rapidly than relaxation of the entire cluster \citep{gerhard00}.  The
near-IR light is dominated by RSG stars which are evolved from
high-mass main sequence stars --- stars which are expected to approach
energy equipartition most rapidly.  In addition, there is evidence of
primordial mass segregation in young massive clusters such as the
Orion Nebula Cluster \citep{lynne98}, R136 in 30 Doradus
\citep[e.g.,][]{brandl96}, and several clusters in the Large
Magellanic Cloud \citep{degrijs02c}.  If star formation processes have
conspired to concentrate high mass stars at the center of the
clusters, the relaxation timescale there will be shorter; dynamical
mass segregation will proceed more rapidly at the core of the
clusters, on the order of a few crossing times \citep{degrijs02b}.  As
a result, the cores of SSCs may undergo significant dynamical
evolution in as little as 25 Myr \citep{degrijs02a}, somewhat younger
than MGG-F.

It is therefore not unlikely that there is significant mass
segregation, either primordial or dynamical, in the clusters.  In that
case, the RSG starlight observed in F160W samples only the inner
portion of the clusters which is dominated by evolved high-mass stars.
The velocity dispersions determined by our cross-correlation analysis
are affected by only the mass located within the volume occupied by
RSG stars.  If a significant amount of mass lies at larger radii, the
half-light radius measured for the RSG stars will not trace the
half-mass radius, leading to an underestimation of the cluster mass.

If, for example, the half-mass radius of cluster MGG-11 is greater
than the measured RSG half-light radius by a factor of 1.5, the
$L_{F160}/M$ ratio would be in line with a standard Kroupa IMF.  Under
such a segregation of masses, the RSG stars would occupy a volume
filling factor of 30\% within the cluster.

\subsection{Fate of the Clusters}
\label{fate}

Whether SSCs MGG-9 and MGG-11 will resemble globular clusters 10--15
billion years from now depends in large part upon two issues: whether
the clusters can remain bound despite mass loss due to stellar
evolution and whether that evolution leads to a mass function simliar
to old globular clusters.  We will deal with these issues in turn.

With ages in excess of 7 Myr, the OB stars formed in the clusters have
begun evolving off the main sequence.  We see no spectral evidence of
nebular emission; stellar winds and ionization from the massive stars
have cleared residual gas from the clusters' formation.  In addition,
any O stars in the clusters have likely exploded as supernovae,
clearing any gas not removed gradually during the main sequence
lifetime of the OB stars.  Studies \citep[e.g.,][]{hills80,goodwin97}
have shown that loss of more than half of a cluster's mass in less
than a crossing time can dissociate the cluster.\footnote{However,
stars in the low velocity tail of the distribution function can remain
bound in spite of loss of as much as 80\% of the cluster mass
\citep[e.g.,][]{adams00}.}  The context for such studies is expulsion
of residual gas and the implied star formation efficiencies of bound
clusters.  Although the residual gas has been removed from MGG-9 and
MGG-11, the clusters will continue to lose mass over time due to
evolution of the member stars.  If the mass loss is gradual, the
stellar orbits have time to adjust to the new potential and the
cluster will simply expand to a new equilibrium radius.  The remaining
opportunity for substantial, impulsive gas loss is from supernovae of
B stars.  For a standard Kroupa IMF, 14\% of the cluster's mass is in
stars with mass $\ge 8$M$_{\odot}$.  Truncating the lower end of the
IMF at 1M$_{\odot}$ shifts more of the cluster's mass to these
supernova progenitors: 35\% would be in stars with mass $\ge
8$M$_{\odot}$.  In either case, however, there is not enough mass for
impulsive mass loss via supernovae to dissociate the clusters.

While the clusters are likely to remain bound in spite of internal
threats, they still must face various external destruction mechanisms
(e.g., evaporation, disk-shocking and dynamical friction).  Although
the prospects of these clusters against these mechanisms is beyond the
scope of the present work, we can make some general comments.
Low-mass clusters are more easily dispersed by each of these processes
\citep{ashman98}; the large masses of MGG-9 and MGG-11 (relative to
Galactic globular clusters) will make these clusters less prone to
disruption.  The low galactocentric radii of the clusters, however,
will tend to make them more prone to disruption by dynamical friction.

The second issue is whether the SSCs will resemble globular clusters
after 10--15 Gyr of stellar evolution.  The most metal-poor globular
clusters in the Galaxy are about 15 Gyr old \citep{vandenberg96}, with
little or no dispersion in age among the Galactic globular cluster
system \citep{stetson96}.  These old globular clusters have main
sequence turn-offs at $\sim 0.8$ M$_{\odot}$, with higher-mass stars
having evolved, leaving behind remnants.  \citet{sternberg98} quotes
typical mass fractions for the component objects in globular clusters:
$f($MS$) \sim 0.73$, $f($WD$) \sim 0.25$, and $f($NS$) \sim 0.02$ for
the fraction of the mass contained in main sequence stars, white
dwarfs and neutron stars, respectively.  Applying Sternberg's simple
model for mapping main sequence progenitors to remnants, we are able
to estimate the composition of a cluster after 15 Gyr of evolution
(assuming the cluster remains bound and loses no stars, e.g. due to
evaporation).

For the case of a power law IMF truncated at a lower mass of 1
M$_{\odot}$, it is clear that no stars will remain on the main
sequence after 15 Gyr; such a cluster could not resemble a globular
cluster as described here.  More interesting are the effects of
evolution on the Kroupa IMF of Section \ref{m2l}.  After 15 Gyr, the
cluster will have lost 51\% of its SSC mass (i.e., the mass present at
an age of 10 Myr).  The composition of the cluster would be $f($MS$) =
0.60$, $f($WD$)=0.37$, and $f($NS$) = 0.04$.  This is similar to the
typical globular cluster composition quoted by Sternberg, but with a
larger proportion of the mass in the form of stellar remnants.
Evolving MGG-9 for a globular cluster lifetime of 15 Gyr leaves a
cluster of M$ = 8 \times 10^5$ M$_{\odot}$ with a composition somewhat
high in remnants but not dissimilar to Galactic globular clusters.

Finally, we note that low-mass stars are preferentially lost in the
process of stellar evaporation.  This effect is exacerbated by mass
segregation, as low-mass stars are in a shallower potential and have
lower escape velocities.  Dynamical effects are likely to decrease the
mass fraction of low-mass main sequence stars, particularly in
low-mass clusters for which these effects are strongest.

\section{Conclusions}
\label{conc}

We have used high resolution near-IR spectra to measure the velocity
dispersions of two spatially-resolved super star clusters in the
nuclear starburst of M82.  We have combined these data with
high-resolution NICMOS images to determine the kinematic masses and
light-to-mass ratios of these SSCs.  The near-IR spectra of the
clusters are dominated by the light of RSG stars, consistent with ages
of 7--12 Myr.

The derived masses, sizes and ages are consistent with the hypothesis
that the SSCs are young globular clusters.  Whether they can evolve
into objects resembling old Galactic globular clusters depends upon
the IMF of the clusters.  When compared to population synthesis
models, the light-to-mass ratio of MGG-9 appears to be consistent with
a standard Kroupa IMF, and the cluster could resemble a globular
cluster 15 Gyr from now.  The light-to-mass ratios of MGG-11 and MGG-F
suggest a deficiency of low-mass stars; it is unlikely that the
composition of these clusters will resemble globular clusters after 15
Gyr of stellar evolution.  There is evidence for mass segregation in
MGG-F; if the red supergiant stars are sampling only the inner portion
of the clusters, our mass estimates would represent lower limits.
This could account for the apparent deficiency of low-mass stars and
bring the $L_{F160}/M$ ratios in line with standard IMFs.  The
presence of mass segregation would further suggest that the SSCs are
scaled-up, high-mass analogues of young Galactic clusters (e.g., the
Orion Nebula Cluster and NGC 3603).

These results are critically dependent on extinction estimates.  The
population of clusters in the nuclear starburst suffers from a wide
range of differential extinction, complicating optical studies.  The
lower extinction in the near-infrared is a distinct advantage in
determination of cluster properties in young, dusty starburst regions.

In order to assess whether the young SSCs in M82 can evolve into a
population of old globular clusters, they should be characterized both
as a system and as individual objects.  The present work represents a
pilot project in preparation for a more complete study of the SSCs in
the nuclear starburst.  We have undertaken an observing program at the
Keck Observatory to use NIRSPEC to measure velocity dispersions for
additional SSCs in the nuclear starburst.  With these spectra, we will
determine the masses and light-to-mass ratios for the population of
SSCs and examine the luminosity and mass functions of the cluster
system.  The presence of system-wide deviations from a standard IMF
would be more compelling than suggestive results for individual
clusters.

%%%%%%%%%%%%%%%%%%%
% Acknowledgments %
%%%%%%%%%%%%%%%%%%%

\acknowledgments

We would like to thank the staff of the Keck Observatory, and
observing assistant Ron Quick in particular.  The authors wish to
recognize and acknowledge the very significant cultural role and
reverence that the summit of Mauna Kea has always had within the
indigenous Hawaiian community.  We are most fortunate to have the
opportunity to conduct observations from this mountain.  We also thank
the anonymous referee for helpful guidance.  NM thanks John Garrett
for inspiration and encouragement to enter science, the Radio
Astronomy Lab at UC Berkeley for financial support, J.T. Wright for
mathematical assistance, and J. Hall for help with spectral line
identifications.  This work has been supported in part by the National
Science Foundation Science and Technology Center for Adaptive Optics,
managed by the University of California at Santa Cruz under
cooperative agreement No. AST-9876783.

%%%%%%%%%%%%%%
% References %
%%%%%%%%%%%%%%

%\eject

\bibliographystyle{apj}
\bibliography{apj-jour,astro_refs}

\begin{thebibliography}{55}
\expandafter\ifx\csname natexlab\endcsname\relax\def\natexlab#1{#1}\fi

\bibitem[{{Adams}(2000)}]{adams00}
{Adams}, F.~C. 2000, \apj, 542, 964

\bibitem[{{Appleton} {et~al.}(1981){Appleton}, {Davies}, \&
  {Stephenson}}]{appleton81}
{Appleton}, P.~N., {Davies}, R.~D., \& {Stephenson}, R.~J. 1981, \mnras, 195,
  327

\bibitem[{{Ashman} \& {Zepf}(1998)}]{ashman98}
{Ashman}, K.~M. \& {Zepf}, S.~E. 1998, {Globular Cluster Systems} (New York :
  Cambridge University Press)

\bibitem[{{Bonnell} \& {Davies}(1998)}]{bonnell98}
{Bonnell}, I.~A. \& {Davies}, M.~B. 1998, \mnras, 295, 691

\bibitem[{{Brandl} {et~al.}(1996){Brandl}, {Sams}, {Bertoldi}, {Eckart},
  {Genzel}, {Drapatz}, {Hofmann}, {Loewe}, \& {Quirrenbach}}]{brandl96}
{Brandl}, B., {Sams}, B.~J., {Bertoldi}, F., {Eckart}, A., {Genzel}, R.,
  {Drapatz}, S., {Hofmann}, R., {Loewe}, M., \& {Quirrenbach}, A. 1996, \apj,
  466, 254

\bibitem[{{Cardelli} {et~al.}(1989){Cardelli}, {Clayton}, \&
  {Mathis}}]{cardelli89}
{Cardelli}, J.~A., {Clayton}, G.~C., \& {Mathis}, J.~S. 1989, \apj, 345, 245

\bibitem[{{de Grijs} {et~al.}(2002{\natexlab{a}}){de Grijs}, {Gilmore},
  {Johnson}, \& {Mackey}}]{degrijs02b}
{de Grijs}, R., {Gilmore}, G.~F., {Johnson}, R.~A., \& {Mackey}, A.~D.
  2002{\natexlab{a}}, \mnras, 331, 245

\bibitem[{{de Grijs} {et~al.}(2002{\natexlab{b}}){de Grijs}, {Gilmore},
  {Mackey}, {Wilkinson}, {Beaulieu}, {Johnson}, \& {Santiago}}]{degrijs02c}
{de Grijs}, R., {Gilmore}, G.~F., {Mackey}, A.~D., {Wilkinson}, M.~I.,
  {Beaulieu}, S.~F., {Johnson}, R.~A., \& {Santiago}, B.~X. 2002{\natexlab{b}},
  \mnras, 337, 597

\bibitem[{{de Grijs} {et~al.}(2002{\natexlab{c}}){de Grijs}, {Johnson},
  {Gilmore}, \& {Frayn}}]{degrijs02a}
{de Grijs}, R., {Johnson}, R.~A., {Gilmore}, G.~F., \& {Frayn}, C.~M.
  2002{\natexlab{c}}, \mnras, 331, 228

\bibitem[{{de~Grijs} {et~al.}(2001){de~Grijs}, {O'Connell}, \&
  {Gallagher}}]{degrijs01}
{de~Grijs}, R., {O'Connell}, R.~W., \& {Gallagher}, J.~S. 2001, \aj, 121, 768

\bibitem[{{Eisenhauer}(2001)}]{eisenhauer01}
{Eisenhauer}, F. 2001, in Starburst Galaxies: Near and Far, ed. L.~Tacconi \&
  D.~Lutz (Berlin: Springer), 24

\bibitem[{{F\"{o}rster Schreiber}(1998)}]{natascha98}
{F\"{o}rster Schreiber}, N.~M. 1998, PhD thesis,
  Lugwig-Maximilians-Universit\"{a}t M\"{u}nchen

\bibitem[{{F\"{o}rster Schreiber} {et~al.}(2001){F\"{o}rster Schreiber},
  {Genzel}, {Lutz}, {Kunze}, \& {Sternberg}}]{natascha01}
{F\"{o}rster Schreiber}, N.~M., {Genzel}, R., {Lutz}, D., {Kunze}, D., \&
  {Sternberg}, A. 2001, \apj, 552, 544

\bibitem[{{Freedman} {et~al.}(1994){Freedman}, {Hughes}, {Madore}, {Mould},
  {Lee}, {Stetson}, {Kennicutt}, {Turner}, {Ferrarese}, {Ford}, {Graham},
  {Hill}, {Hoessel}, {Huchra}, \& {Illingworth}}]{freedman94}
{Freedman}, W.~L., {Hughes}, S.~M., {Madore}, B.~F., {Mould}, J.~R., {Lee},
  M.~G., {Stetson}, P., {Kennicutt}, R.~C., {Turner}, A., {Ferrarese}, L.,
  {Ford}, H., {Graham}, J.~A., {Hill}, R., {Hoessel}, J.~G., {Huchra}, J., \&
  {Illingworth}, G.~D. 1994, \apj, 427, 628

\bibitem[{{Gerhard}(2000)}]{gerhard00}
{Gerhard}, O. 2000, in ASP Conf. Ser. 211: Massive Stellar Clusters, ed.
  A.~Lan\c{c}on \& C.~M. Boily (San Francisco: ASP), 12

\bibitem[{{Gilbert}(2002)}]{gilbert02th}
{Gilbert}, A.~M. 2002, PhD thesis, Univ. of California, Berkeley

\bibitem[{{Gilbert} \& {Graham}(2001)}]{gilbert01}
{Gilbert}, A.~M. \& {Graham}, J.~R. 2001, in Starburst Galaxies: Near and Far,
  ed. L.~Tacconi \& D.~Lutz (Berlin: Springer), 123

\bibitem[{{Gilbert} \& {Graham}(2002)}]{gilbert02iau}
{Gilbert}, A.~M. \& {Graham}, J.~R. 2002, in IAU Symposium 207: Extragalactic
  Star Clusters, ed. D.~Geisler, E.~K. Grebel, \& D.~Minniti (San Francisco:
  ASP), 471

\bibitem[{{Gilmore}(2001)}]{gilmore01}
{Gilmore}, G. 2001, in Starburst Galaxies: Near and Far, ed. L.~Tacconi \&
  D.~Lutz (Berlin: Springer), 34

\bibitem[{{Goodwin}(1997)}]{goodwin97}
{Goodwin}, S.~P. 1997, \mnras, 284, 785

\bibitem[{{Gray} \& {Toner}(1987)}]{gray87}
{Gray}, D.~F. \& {Toner}, C.~G. 1987, \apj, 322, 360

\bibitem[{{Heckman}(1998)}]{heckman98}
{Heckman}, T.~M. 1998, in ASP Conf. Ser. 148: Origins, ed. C.~E. Woodward,
  J.~M. Shull, \& H.~A. Thronson (San Francisco: ASP), 127

\bibitem[{{Hillenbrand} \& {Hartmann}(1998)}]{lynne98}
{Hillenbrand}, L.~A. \& {Hartmann}, L.~W. 1998, \apj, 492, 540

\bibitem[{{Hills}(1980)}]{hills80}
{Hills}, J.~G. 1980, \apj, 235, 986

\bibitem[{{Ho} \& {Filippenko}(1996{\natexlab{a}})}]{hoflipper96a}
{Ho}, L.~C. \& {Filippenko}, A.~V. 1996{\natexlab{a}}, \apjl, 466, L83

\bibitem[{{Ho} \& {Filippenko}(1996{\natexlab{b}})}]{hoflipper96b}
---. 1996{\natexlab{b}}, \apj, 472, 600

\bibitem[{{King}(1962)}]{king62}
{King}, I. 1962, \aj, 67, 471

\bibitem[{{Krist}(1995)}]{Krist95}
{Krist}, J. 1995, in ASP Conf. Ser. 77: Astronomical Data Analysis Software and
  Systems IV, ed. R.~Shaw, H.~E. Payne, \& J.~E. Hayes, Vol.~4 (San Francisco:
  ASP), 349

\bibitem[{{Kronberg} {et~al.}(1972){Kronberg}, {Pritchet}, \& {van den
  Bergh}}]{kronberg72}
{Kronberg}, P.~P., {Pritchet}, C.~J., \& {van den Bergh}, S. 1972, \apjl, 173,
  L47

\bibitem[{{Kroupa}(2001)}]{kroupa01}
{Kroupa}, P. 2001, \mnras, 322, 231

\bibitem[{{Lan{\c c}on} {et~al.}(1999){Lan{\c c}on}, {Mouhcine}, {Fioc}, \&
  {Silva}}]{lancon99}
{Lan{\c c}on}, A., {Mouhcine}, M., {Fioc}, M., \& {Silva}, D. 1999, \aap, 344,
  L21

\bibitem[{{Leitherer} {et~al.}(1999){Leitherer}, {Schaerer}, {Goldader},
  {Delgado}, {Robert}, {Kune}, {de Mello}, {Devost}, \&
  {Heckman}}]{leitherer99}
{Leitherer}, C., {Schaerer}, D., {Goldader}, J.~D., {Delgado}, R.~M.~G.~.,
  {Robert}, C., {Kune}, D.~F., {de Mello}, D.~.~F., {Devost}, D., \& {Heckman},
  T.~M. 1999, \apjs, 123, 3

\bibitem[{{Marleau} {et~al.}(2000){Marleau}, {Graham}, {Liu}, \&
  {Charlot}}]{marleau00}
{Marleau}, F.~R., {Graham}, J.~R., {Liu}, M.~C., \& {Charlot}, S. 2000, \aj,
  120, 1779

\bibitem[{{McLean} {et~al.}(1998){McLean}, {Becklin}, {Bendiksen}, {Brims},
  {Canfield}, {Figer}, {Graham}, {Hare}, {Lacayanga}, {Larkin}, {Larson},
  {Levenson}, {Magnone}, {Teplitz}, \& {Wong}}]{mclean98}
{McLean}, I.~S., {Becklin}, E.~E., {Bendiksen}, O., {Brims}, G., {Canfield},
  J., {Figer}, D.~F., {Graham}, J.~R., {Hare}, J., {Lacayanga}, F., {Larkin},
  J.~E., {Larson}, S.~B., {Levenson}, N., {Magnone}, N., {Teplitz}, H., \&
  {Wong}, W. 1998, in Proc. SPIE Vol. 3354, Infrared Astronomical
  Instrumentation, ed. A.~M. Fowler (Bellingham: SPIE), 566

\bibitem[{{McLeod} {et~al.}(1993){McLeod}, {Rieke}, {Rieke}, \&
  {Kelly}}]{mcleod93}
{McLeod}, K.~K., {Rieke}, G.~H., {Rieke}, M.~J., \& {Kelly}, D.~M. 1993, \apj,
  412, 111

\bibitem[{{Mengel} {et~al.}(2002){Mengel}, {Lehnert}, {Thatte}, \&
  {Genzel}}]{mengel02}
{Mengel}, S., {Lehnert}, M.~D., {Thatte}, N., \& {Genzel}, R. 2002, \aap, 383,
  137

\bibitem[{{Meurer} {et~al.}(1995){Meurer}, {Heckman}, {Leitherer}, {Kinney},
  {Robert}, \& {Garnett}}]{meurer95}
{Meurer}, G.~R., {Heckman}, T.~M., {Leitherer}, C., {Kinney}, A., {Robert}, C.,
  \& {Garnett}, D.~R. 1995, \aj, 110, 2665

\bibitem[{{Meyer} {et~al.}(2000){Meyer}, {Adams}, {Hillenbrand}, {Carpenter},
  \& {Larson}}]{meyer00}
{Meyer}, M.~R., {Adams}, F.~C., {Hillenbrand}, L.~A., {Carpenter}, J.~M., \&
  {Larson}, R.~B. 2000, in Protostars and Planets IV, ed. V.~Mannings, A.~P.
  Boss, \& S.~S. Russell (Tucson: Univ. Arizona Press), 121

\bibitem[{{Meylan}(1987)}]{meylan87}
{Meylan}, G. 1987, \aap, 184, 144

\bibitem[{{Miller} \& {Scalo}(1978)}]{miller78}
{Miller}, G.~E. \& {Scalo}, J.~M. 1978, \pasp, 90, 506

\bibitem[{{Mouhcine} \& {Lan{\c c}on}(2002)}]{mouhcine02}
{Mouhcine}, M. \& {Lan{\c c}on}, A. 2002, \aap, 393, 149

\bibitem[{{O'Connell} {et~al.}(1995){O'Connell}, {Gallagher}, {Hunter}, \&
  {Colley}}]{o'connell95}
{O'Connell}, R.~W., {Gallagher}, J.~S., {Hunter}, D.~A., \& {Colley}, W.~N.
  1995, \apjl, 446, L1

\bibitem[{{Pickles}(1998)}]{pickles98}
{Pickles}, A.~J. 1998, \pasp, 110, 863

\bibitem[{{Rieke} {et~al.}(1993){Rieke}, {Loken}, {Rieke}, \&
  {Tamblyn}}]{rieke93}
{Rieke}, G.~H., {Loken}, K., {Rieke}, M.~J., \& {Tamblyn}, P. 1993, \apj, 412,
  99

\bibitem[{{Salpeter}(1955)}]{salpeter55}
{Salpeter}, E.~E. 1955, \apj, 121, 161

\bibitem[{{Satyapal} {et~al.}(1997){Satyapal}, {Watson}, {Pipher}, {Forrest},
  {Greenhouse}, {Smith}, {Fischer}, \& {Woodward}}]{satyapal97}
{Satyapal}, S., {Watson}, D.~M., {Pipher}, J.~L., {Forrest}, W.~J.,
  {Greenhouse}, M.~A., {Smith}, H.~A., {Fischer}, J., \& {Woodward}, C.~E.
  1997, \apj, 483, 148

\bibitem[{{Shen} \& {Lo}(1995)}]{shenlo95}
{Shen}, J. \& {Lo}, K.~Y. 1995, \apjl, 445, L99

\bibitem[{{Smith} \& {Gallagher}(2001)}]{smith01}
{Smith}, L.~J. \& {Gallagher}, J.~S. 2001, \mnras, 326, 1027

\bibitem[{{Spitzer}(1987)}]{spitzer87}
{Spitzer}, L. 1987, {Dynamical Evolution of Globular Clusters} (Princeton, NJ,
  Princeton University Press, 1987)

\bibitem[{{Spitzer}(1969)}]{spitzer69}
{Spitzer}, L.~J. 1969, \apjl, 158, L139

\bibitem[{{Sternberg}(1998)}]{sternberg98}
{Sternberg}, A. 1998, \apj, 506, 721

\bibitem[{{Stetson} {et~al.}(1996){Stetson}, {Vandenberg}, \&
  {Bolte}}]{stetson96}
{Stetson}, P.~B., {Vandenberg}, D.~A., \& {Bolte}, M. 1996, \pasp, 108, 560

\bibitem[{{Tonry} \& {Davis}(1979)}]{tonry79}
{Tonry}, J. \& {Davis}, M. 1979, \aj, 84, 1511

\bibitem[{{Vandenberg} {et~al.}(1996){Vandenberg}, {Stetson}, \&
  {Bolte}}]{vandenberg96}
{Vandenberg}, D.~A., {Stetson}, P.~B., \& {Bolte}, M. 1996, \araa, 34, 461

\bibitem[{{Zepf} {et~al.}(1999){Zepf}, {Ashman}, {English}, {Freeman}, \&
  {Sharples}}]{zepf99}
{Zepf}, S.~E., {Ashman}, K.~M., {English}, J., {Freeman}, K.~C., \& {Sharples},
  R.~M. 1999, \aj, 118, 752

\end{thebibliography}

%%%%%%%%%%%
% Figures %
%%%%%%%%%%%

\clearpage

%-----------%
% HST Image %
%-----------%

%------------%
% IR Spectra %
%------------%

%--------------%
% X-Corr Plots %
%--------------%

%------------------%
% Radial Profile   %
%------------------%

%-------------------%
% H vs I Comparison %
%-------------------%

%-------------------------%
% Color-Magnitude Diagram %
%-------------------------%

%------------------%
% Starburst99 plot %
%------------------%

%%%%%%%%%%
% Tables %
%%%%%%%%%%

\clearpage

%----------------------------%
% Template Star Observations %
%----------------------------%

\begin{deluxetable}{lcccc}
\tablewidth{0pt}
\tablecolumns{5}
\tablecaption{Template Star Observations}
\tablehead{\colhead{Template Star} & \colhead{Spectral Type} & \colhead{Date} & \colhead{Airmass} & \colhead{Atmospheric Calibrator} }
\startdata
HR 3459   & G2Iab & 2000 Dec 10 & 1.12 & HR 2432 (B1Ib) \\
HD 186293 & K0I   & 2001 Jun 17 & 1.08 & HR 7767 (O9V) \\
HR 2119   & K0II  & 2000 Dec 10 & 1.26 & HR 2432 (B1Ib) \\
HR 8248   & K1Iab & 2001 Jun 17 & 1.11 & HR 7767 (O9V) \\
HD 38054  & K3III & 2002 Feb 23 & 1.53 & HD 32249 (B3V) \\
HD 201251 & K4Iab & 2001 Jun 17 & 1.14 & HR 7767 (O9V) \\
HD 185622 & K4Ib  & 2001 Jun 17 & 1.08 & HR 7767 (O9V) \\
HD 200905 & K4.5Ib& 2001 Jun 17 & 1.10 & HR 7767 (O9V) \\ 
HD 43335  & K5Iab & 2002 Feb 23 & 1.18 & HD 41753 (B3V) \\ 
HR 2289   & M0Iab & 2000 Dec 10 & 1.28 & HR 2432 (B1Ib) \\
HD 14488  & M4.5Iab& 2003 Jan 19 & 1.26 & HD 14372 (B5V)
\enddata
\label{calstars}
\end{deluxetable}

%--------------------%
% Cross Correlations %
%--------------------%

\begin{deluxetable}{lccccc}
\tablewidth{0pt}
\tablecolumns{6}
\tablecaption{Cross-Correlation Results}
\tablehead{\colhead{} & \colhead{} & \multicolumn{2}{c}{MGG-9} & \multicolumn{2}{c}{MGG-11} \\ 
\colhead{Template} & \colhead{Spectral} & \colhead{Cross-Correlation} & \colhead{$\sigma_{CCF}$\tablenotemark{b}} & \colhead{Cross-Correlation} & \colhead{$\sigma_{CCF}$\tablenotemark{b}} \\
\colhead{Star} & \colhead{Type} & \colhead{Peak\tablenotemark{a}} & \colhead{(km s$^{-1}$)} &
\colhead{Peak\tablenotemark{a}} & \colhead{(km s$^{-1}$)} }
\startdata
HR 3459   & G2Iab & $0.33 \pm 0.02$ & $17   \pm 2  $ & $0.33 \pm 0.03$ & $11   \pm 2  $ \\
HD 186293 & K0I   & $0.42 \pm 0.02$ & $19   \pm 1.2  $ & $0.47 \pm 0.02$ & $11   \pm 2  $ \\
HR 2119   & K0II  & $0.39 \pm 0.02$ & $20   \pm 1.1  $ & $0.44 \pm 0.02$ & $12  \pm 1.1 $ \\
HR 8248   & K1Iab & $0.62 \pm 0.01$ & $17.2 \pm 0.8$ & $0.69 \pm 0.02$ & $12.0 \pm 0.8$ \\
HD 38054  & K3III & $0.52 \pm 0.01$ & $18.2 \pm 0.8$ & $0.59 \pm 0.02$ & $12.3 \pm 0.8$ \\
HD 201251 & K4Iab & $0.59 \pm 0.01$ & $17.3 \pm 0.8$ & $0.66 \pm 0.02$ & $12.0 \pm 0.8$ \\
HD 185622 & K4Ib  & $0.68 \pm 0.01$ & $16.8 \pm 0.8$ & $0.75 \pm 0.01$ & $11.9 \pm 0.8$ \\
HD 200905 & K4.5Ib & $0.64 \pm 0.01$ & $17.0 \pm 0.8$ & $0.72 \pm 0.01$ & $12.1 \pm 0.8$ \\ 
HD 43335  & K5Iab & $0.58 \pm 0.01$ & $17.5 \pm 0.7$ & $0.67 \pm 0.01$ & $12.4 \pm 0.8$ \\ 
HR 2289   & M0Iab & $0.72 \pm 0.01$ & $16.2 \pm 0.8$ & $0.79 \pm 0.01$ & $11.3 \pm 0.7$ \\
HD 14488  & M4.5Iab& $0.72 \pm 0.01$ & $15.5 \pm 0.7$ & $0.79 \pm 0.01$ & $11.5 \pm 0.8$
\enddata
\tablenotetext{a}{Maximum value of the cross-correlation function.}
\tablenotetext{b}{$\sigma_{CCF}$ is the stellar velocity dispersion as
determined from the cross-correlation function.}
\label{xcorr}
\end{deluxetable}

%---------------------%
% Cluster Fit Results %
%---------------------%

\begin{deluxetable}{cccccccc}
%\tabletypesize{\scriptsize}
\tablewidth{0pt}
\tablecolumns{8}
\tablecaption{Super Star Cluster Parameters}
\tablehead{\colhead{SSC} & \colhead{MGG} & \colhead{R.A.\tablenotemark{\dagger}} & \colhead{Dec.\tablenotemark{\dagger}} & 
\colhead{$\left[F160W\right]$\tablenotemark{\ast}} & \colhead{$\left[F222M\right]$\tablenotemark{\ast}} & 
\colhead{$r_{hp}$\tablenotemark{\ddagger}}  \\
\colhead{ } & \colhead{} & \colhead{(sec)} & \colhead{(arcsec)} & \colhead{(mags)} & \colhead{(mags) } & \colhead{(mas)}}
\startdata
J0955468+694937  &  L  &  46.8  &  37.1  &  12.05 $\pm$  0.03  &  11.45 $\pm$  0.03  &  83 $\pm$ 10  \\
J0955471+694940  &  F  &  47.1  &  40.8  &  12.96 $\pm$  0.04  &  12.65 $\pm$  0.04  &  89 $\pm$ 11  \\
J0955483+694941  &  b  &  48.3  &  41.0  &  14.5  $\pm$  0.1   &  13.46 $\pm$  0.06  & 113 $\pm$ 14  \\
J0955488+694942  &  c  &  48.8  &  42.5  &  16.1  $\pm$  0.4   &  14.7  $\pm$  0.5   &  73 $\pm$  9  \\
J0955489+694940  &  d  &  48.9  &  40.5  &  15.5  $\pm$  0.2   &  14.5  $\pm$  0.2   &  94 $\pm$ 11  \\
J0955502+694945  & 11  &  50.2  &  45.1  &  13.10 $\pm$  0.15  &  12.03 $\pm$  0.05  &  66 $\pm$  8  \\
J0955505+694945  &  9  &  50.5  &  45.1  &  12.90 $\pm$  0.15  &  11.55 $\pm$  0.05  & 146 $\pm$ 18  \\
J0955506+694942  &  8  &  50.6  &  42.9  &  14.3  $\pm$  0.3   &  13.13 $\pm$  0.15  &  91 $\pm$ 11  \\
J0955510+694946  &  7  &  51.0  &  46.6  &  13.4  $\pm$  0.2   &  12.80 $\pm$  0.15  & 156 $\pm$ 19  \\
J0955513+694946  &  6  &  51.3  &  46.8  &  13.03 $\pm$  0.15  &  12.18 $\pm$  0.05  &  79 $\pm$  9  \\
J0955525+694943  &  k  &  52.5  &  43.2  &  14.1  $\pm$  0.3   &  13.8  $\pm$  0.3   & 166 $\pm$ 20  \\
J0955526+694943  &  m  &  52.6  &  43.0  &  14.8  $\pm$  0.7   &  14.0  $\pm$  0.3   &  79 $\pm$  9  \\
J0955528+694950  &  q  &  52.8  &  50.6  &  14.3  $\pm$  0.3   &  13.7  $\pm$  0.2   & 113 $\pm$ 14  \\
J0955529+694944  &  3  &  52.9  &  44.4  &  13.3  $\pm$  0.2   &  12.37 $\pm$  0.08  &  90 $\pm$ 11  \\
J0955535+694949  & 1a  &  53.5  &  49.7  &  12.90 $\pm$  0.15  &  12.43 $\pm$  0.08  & 119 $\pm$ 14  \\
J0955536+694948  & 1b  &  53.6  &  48.4  &  13.5  $\pm$  0.2   &  12.8  $\pm$  0.1   & 108 $\pm$ 13  \\
J0955537+694949  & 1c  &  53.8  &  49.4  &  13.5  $\pm$  0.2   &  12.7  $\pm$  0.1   &  87 $\pm$ 11  \\
J0955548+694953  &  r  &  54.8  &  53.9  &  13.43 $\pm$  0.06  &  12.67 $\pm$  0.04  & 101 $\pm$ 12  \\
J0955560+694952  &  s  &  56.0  &  52.3  &  15.25 $\pm$  0.15  &  14.2  $\pm$  0.1   & 105 $\pm$ 13  \\
J0955571+694954  &  t  &  57.1  &  54.0  &  14.93 $\pm$  0.15  &  14.06 $\pm$  0.09  & 100 $\pm$ 12  \\
\enddata
\tablenotetext{\dagger}{Offsets from coordinates: R.A. = 09 55 00.0, Dec. =
69 40 00.0 (J2000.0).  The quoted astrometry is $\pm 0.''6$.}
\tablenotetext{\ast}{Total magnitudes (uncorrected for reddening)
based upon integration of King model fits to the data.}
\tablenotetext{\ddagger}{The projected half-light radius, $r_{hp}$, is
based on King model fits to the F160W image.}
\tablecomments{As recommended by the IAU, we have assigned each SSC a
principal, coordinate-based designation.  In addition, we refer to
each source by a shorter nickname, prefaced by an acronym composed of
the authors' initials, MGG.  }
\label{sscdata}
\end{deluxetable}

\end{document}